\documentstyle[12pt,epsf]{article}
\thicklines
\begin{document}
\hfill DESY 97-204
\begin{center}
{\bf\sc Perturbative Fragmentation of Leptoquark into Heavy Lepto-Quarkonium}\\
\vspace*{4mm}
V.V.Kiselev\\

\vspace*{3mm}
Institute for High Energy Physics,\\
Protvino, Moscow Region, 142284, Russia.\\
kiselev@mx.ihep.su
\end{center}

\vspace*{2mm}
\begin{abstract}
The fragmentation function of a scalar leptoquark into possible
S-wave bound states 
with a heavy anti-quark is calculated to the leading order 
in perturbative QCD for the high energy processes at large transverse momenta.
The one-loop equations for the $q^2$-evolution of moments of the 
fragmentation function due to the hard gluon emission by the leptoquark
are derived. The integral probabilities of fragmentation are evaluated.
\end{abstract}

\section{Introduction}
An abundant discussion about a possible appearance of new physics at HERA 
\cite{H1,Zeus,J} has provided an important and useful experience for a
handling of almost-real properties attributed to interactions 
beyond the Standard Model. So, if a leptoquark
\cite{Buch} is the correct interpretation 
for the excess in the HERA data at high $Q^2$, $x$, then the former has
the total width $\Gamma$, which is much less than the QCD-confinement scale,
$\Gamma_{LQ}\ll \Lambda_{QCD}$ \cite{Alta}. The latter fact means that
before a decay, the color-triplet leptoquark LQ can be bound with quarks 
in the lepto-hadrons: $(\bar q LQ)$-baryons or $(q_1q_2LQ)$-mesons, which are
quite exotic states as well as the double LQ-onia: $(\overline{LQ}_1 LQ_2)$.

It is an attractive and interesting problem to study the spectroscopy, 
production and decays of such hadrons as a possible window of new physics
independently of a success or fall off the treatment of the HERA events.

The description of leptoquarks bound with light quarks is a subject of
Leptoquark Effective Theory, which can be developed as a straightforward
continuation of analogous Heavy Quark Effective Theory \cite{Neu}
with taking a care on the spin structure of leptoquark.
The heavy lepto-quarkonia: $(\bar b LQ)$ and $(\bar c LQ)$, can be considered
in the framework of Non-Relativistic QCD \cite{BBL} keeping in mind again the
spin features.

In this work we discuss the high energy production of
heavy lepto-quarkonium containing a scalar leptoquark.

The model-independent pair production of free leptoquarks in hadronic 
collisions was considered in \cite{Z} with account 
for the next-to-leading order QCD corrections. Attaching the result to the
Tevatron search for the scalar leptoquarks \cite{FNAL}, the authors
have found the constraint $m_{LQ}>190$ GeV.

At high transverse momenta, the dominant production mechanism  for the
heavy lepto-quarkonium bound states is the leptoquark fragmentation, 
which can be calculated in perturbative QCD \cite{Bra-rev} after 
the isolation of soft-binding factor extracted from the non-relativistic 
potential models \cite{Martin,Buchm}. The corresponding fragmentation
function is universal for any high energy process for the direct
production of lepto-quarkonia.

In the leading  $\alpha_s$-order, the fragmentation function has 
a scaling form, which is the initial one for the perturbative QCD evolution 
caused by the emission of hard gluons by the leptoquark before 
the hadronization. The corresponding splitting function differs from that for
the heavy quark because of the spin structure of gluon coupling to the
leptoquark.

In this work, the LO-scaling fragmentation function is calculated in Section 2.
The limit of infinitely heavy leptoquark, $m_{LQ}\to \infty$, is obtained
from the full QCD consideration for the fragmentation. The splitting kernel
of the DGLAP-evolution is derived in Section 3, where the one-loop equations
of renormalization group for the moments of fragmentation function are 
obtained and solved. The mentioned equations are universal, since they 
do not depend on whether the leptoquark will bound or free at low virtualities,
where the perturbative evolution stops. The integrated probabilities
of leptoquark fragmentation into the heavy lepto-quarkonia are evaluated in 
Section 4 with making the use of non-relativistic wave-functions 
for the bound states. The results are summarized in Conclusion.

\section{Fragmentation function in leading order}

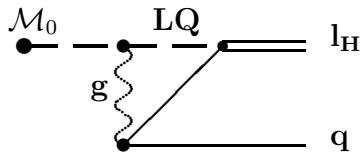
\begin{figure}[th]
\vspace*{25mm}
\begin{center}
\begin{picture}(80,40)
\setlength{\unitlength}{1.1mm}
%\setlinethickness{thick}
\put(2,30){\circle*{2}}
\put(0,32){${\cal M}_0$}
\put(14,30){\circle*{1.5}}
\put(2,30){\line(1,0){4}}
\put(8,30){\line(1,0){4}}
\put(14,30){\line(1,0){4}}
\put(20,30){\line(1,0){4}}
\put(26,30){\circle*{1.3}}
\put(26,30.5){\line(1,0){10}}
\put(26,29.5){\line(1,0){10}}
\put(17.5,32){\bf LQ}
\put(39,30){$\bf l_H$}
\put(14,18){\circle*{1.5}}

\put(14,18){\line(1,1){12}}
\put(14,18){\line(1,0){22}}
\put(39,18){\bf q}
\put(10,24){\bf g}
%%%%%%%%%%%%%%%%%%%%%%%%%%%%
 \put( 13.63,  18.12){\circle*{0.3}}
 \put( 13.32,  18.36){\circle*{0.3}}
 \put( 13.10,  18.60){\circle*{0.3}}
 \put( 13.00,  18.84){\circle*{0.3}}
 \put( 13.05,  19.08){\circle*{0.3}}
 \put( 13.23,  19.32){\circle*{0.3}}
 \put( 13.52,  19.56){\circle*{0.3}}
 \put( 13.87,  19.80){\circle*{0.3}}
 \put( 14.25,  20.04){\circle*{0.3}}
 \put( 14.59,  20.28){\circle*{0.3}}
 \put( 14.84,  20.52){\circle*{0.3}}
 \put( 14.98,  20.76){\circle*{0.3}}
 \put( 14.98,  21.00){\circle*{0.3}}
 \put( 14.84,  21.24){\circle*{0.3}}
 \put( 14.59,  21.48){\circle*{0.3}}
 \put( 14.25,  21.72){\circle*{0.3}}
 \put( 13.87,  21.96){\circle*{0.3}}
 \put( 13.52,  22.20){\circle*{0.3}}
 \put( 13.23,  22.44){\circle*{0.3}}
 \put( 13.05,  22.68){\circle*{0.3}}
 \put( 13.00,  22.92){\circle*{0.3}}
 \put( 13.10,  23.16){\circle*{0.3}}
 \put( 13.32,  23.40){\circle*{0.3}}
 \put( 13.63,  23.64){\circle*{0.3}}
 \put( 14.00,  23.88){\circle*{0.3}}
 \put( 14.37,  24.12){\circle*{0.3}}
 \put( 14.68,  24.36){\circle*{0.3}}
 \put( 14.90,  24.60){\circle*{0.3}}
 \put( 15.00,  24.84){\circle*{0.3}}
 \put( 14.95,  25.08){\circle*{0.3}}
 \put( 14.77,  25.32){\circle*{0.3}}
 \put( 14.48,  25.56){\circle*{0.3}}
 \put( 14.13,  25.80){\circle*{0.3}}
 \put( 13.75,  26.04){\circle*{0.3}}
 \put( 13.41,  26.28){\circle*{0.3}}
 \put( 13.16,  26.52){\circle*{0.3}}
 \put( 13.02,  26.76){\circle*{0.3}}
 \put( 13.02,  27.00){\circle*{0.3}}
 \put( 13.16,  27.24){\circle*{0.3}}
 \put( 13.41,  27.48){\circle*{0.3}}
 \put( 13.75,  27.72){\circle*{0.3}}
 \put( 14.13,  27.96){\circle*{0.3}}
 \put( 14.48,  28.20){\circle*{0.3}}
 \put( 14.77,  28.44){\circle*{0.3}}
 \put( 14.95,  28.68){\circle*{0.3}}
 \put( 15.00,  28.92){\circle*{0.3}}
 \put( 14.90,  29.16){\circle*{0.3}}
 \put( 14.68,  29.40){\circle*{0.3}}
 \put( 14.37,  29.64){\circle*{0.3}}
 \put( 14.00,  29.88){\circle*{0.3}}

%%%%%%%%%%%%%%%%%%%%%
\end{picture}
\end{center}
\vspace*{-18mm}
\caption{The diagram of leptoquark fragmentation 
into the heavy lepto-quarkonium.}
\label{diag}
\end{figure}

The contribution of fragmentation into the direct production of 
heavy lepto-quarkonium has the form
$$
d\sigma[l_H(p)] = \int_0^1 dz\; d\hat \sigma[LQ(p/z),\mu]\; 
D_{LQ\to l_H}(z,\mu),
$$
where $d\sigma$ is the differential cross-section of lepto-quarkonium with the
4-momentum $p$, $d\hat \sigma$ is that of the hard production of
leptoquark with the scaled momentum $p/z$, and $D$ is interpreted as the
fragmentation function depending on the fraction of momentum carried out by
the bound state. The value of $\mu$ determines the factorization scale. 
In accordance with the general DGLAP-evolution, the $\mu$-dependent
fragmentation function satisfies the equation
\begin{equation}
\frac{\partial D_{LQ\to l_H}(z,\mu)}{\partial \ln \mu} =
\int_z^1 \frac{dy}{y} \; P_{LQ\to LQ}(z/y,\mu)\; D_{LQ\to l_H}(y,\mu),
\label{DGLAP}
\end{equation}
where $P$ is the kernel caused by the emission of hard gluons off
the leptoquark leg before the production of heavy quark pair. Therefore, the
initial form of fragmentation function is determined by the diagram 
shown in Fig.\ref{diag}, and, hence, the corresponding initial
factorization scale is equal to $\mu= 2m_Q$. Furthermore,
this function can be calculated as an expansion in $\alpha_s(2m_Q)$.
The leading order contribution is evaluated in this Section.

Consider the fragmentation diagram in the system, where the momentum of 
initial leptoquark has the form $q=(q_0,0,0,q_3)$ and the lepto-quarkonium
one is $p$, so that 
$$
q^2=s, \;\; p^2=M^2.\;\; 
$$
In the static approximation for the bound state of leptoquark and
heavy quark,
the quark mass is expressed as $m_Q= r M$, and the leptoquark mass equals
$m=(1-r)M$.

The matrix element has the form
\begin{equation}
{\cal M} = -\frac{2\sqrt{2\pi}\alpha_s}{3\sqrt{3M^3}}
\frac{R(0)}{r(1-r)(s-m^2)^2} (q^\mu+(1-r)p^\mu) \rho_{\mu\nu}\;
\bar q \gamma^\nu(\hat p-M) l_H \; {\cal M}_0,
\label{M}
\end{equation}
where the sum over the gluon polarizations is written down in the
axial gauge with $n=(1,0,0,-1)$
$$
\rho_{\mu\nu}(k) = -g_{\mu\nu}+\frac{k_\mu n_\nu+k_\nu n_\mu}{k\cdot n},
$$
with $k=q-(1-r)p$. The spinors of $l_H$ and $\bar q$ correspond to
the lepto-quarkonium and heavy quark associated to the fragmentation.
${\cal M}_0$ denotes the matrix element for the hard production of leptoquark
at high energy, $R(0)$ is the radial wave-function at the origin.

Define
$$
z=\frac{p\cdot n}{q\cdot n}
$$
The fragmentation function is determined by the expression \cite{Bra-frag}
$$ 
D(z) = \frac{1}{16\pi^2}\int ds \theta\biggl(s-\frac{M^2}{z}-\frac{m_Q^2}
{1-z}\biggr)\; \frac{|{\cal M}|^2}{|{\cal M}_0|^2},
$$ 
in the limit of high energies $q\cdot n \to \infty$. Then one can 
straightforwardly find
\begin{equation}
D(z) = \frac{8\alpha_s^2}{27\pi}\;
\frac{|R(0)|^2}{M^3 r^2(1-r)^2}\;
\frac{z^2(1-z)^2}{(1-(1-r)z)^6}[(1+r^2)(1+(1-r)^2z^2)-2(1-r)^2(1+r)z],
\label{fra}
\end{equation}
which tends to
\begin{equation}
\tilde D(y)= \frac{8\alpha_s^2}{27\pi}\;
\frac{|R(0)|^2}{m_Q^3}\; \frac{(y-1)^2}{r}\biggl(\frac{4}{y^6}+
\frac{1}{y^4}\biggr),
\label{tilde}
\end{equation}
at $r\to 0$ and $y=(1-(1-r)z)/(rz)$.
The coefficient at the $(y-1)^2$ term is the same as in the fragmentation of
heavy quark with the mass $m$ into the S-wave states of the heavy quarkonium
at $y\to 1$, if one excepts the factor related with 
the wave-function of final state. The limit of $\tilde D(y)$ is in agreement 
with the general consideration of $1/m$-expansion for the fragmentation 
function \cite{JR}, where
$$
\tilde D(y) = \frac{1}{r}a(y) +b(y).
$$
Eq.(\ref{tilde}) determines the $a(y)$-function explicitly.

\section{Hard gluon emission}

The one-loop contribution of hard gluon emission can be calculated in the
way described in previous Section. Then the splitting kernel
of the leptoquark is equal to
\begin{equation}
P_{LQ\to LQ}(x,\mu) = \frac{4\alpha_s(\mu)}{3\pi}\;
\bigg[\frac{2x}{1-x}\biggr]_+,
\label{P}
\end{equation}
where the "plus" denotes the ordinary action: $\int_0^1 dx f_+(x)\cdot g(x)=
\int_0^1 dx f(x)\cdot [g(x)-g(1)]$. The scalar leptoquark splitting
function can be compared with that of the heavy quark
$$
P_{Q\to Q}(x,\mu) = \frac{4\alpha_s(\mu)}{3\pi}\;
\bigg[\frac{1+x^2}{1-x}\biggr]_+,
$$
which has the same normalization factor at $x\to 1$.

Further, multiplying the evolution equation by $z^n$ and integrating over $z$,
one can get from eq.(\ref{DGLAP}) the $\mu$-dependence of moments 
$a_{(n)}$ for the fragmentation function to the one-loop 
accuracy of renormalization group,
\begin{equation}
\frac{\partial a_{(n)}}{\partial \ln \mu} = - \frac{8\alpha_s(\mu)}{3\pi}\;
\bigg[\frac{1}{2}+\ldots +\frac{1}{n+1}\biggr]\; a_{(n)}, \;\;\; n\ge 1.
\label{da}
\end{equation}
At $n=0$ the right hand side of (\ref{da}) equals zero, which means that
the integral probability of leptoquark fragmentation into the heavy
lepto-quarkonium is not changed during the evolution, and it is determined
by the initial fragmentation function calculated perturbatively in
previous Section.

\begin{figure}[th]
\hspace*{3cm}
\epsfxsize=8cm \epsfbox{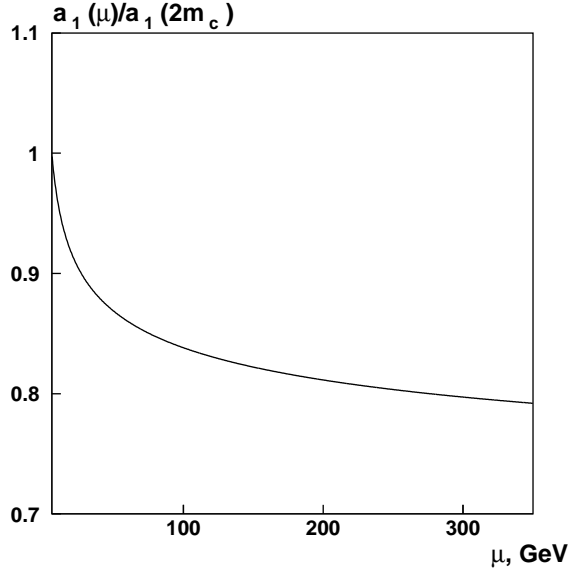}
\caption{The QCD-evolution for the  averaged fraction of scalar leptoquark
momentum, as it is developed in the fragmentation with account for the  
gluon emission to the scale $\mu$, characterizing the hard production of
leptoquark, from the initial value chosen $\mu_0=2m_c$.} 
\label{a1-fig}
\end{figure}

The solution of eq.(\ref{da}) has the form
\begin{equation}
a_{(n)}(\mu) = a_{(n)}(\mu_0)\; \biggl[\frac{\alpha_s(\mu)}
{\alpha_s(\mu_0)}\biggr]^{\frac{16}{3\beta_0}
\bigg[\frac{1}{2}+\ldots +\frac{1}{n+1}\biggr]},
\label{a}
\end{equation}
where one has used the one-loop expression for the QCD coupling constant
$$
\alpha_s(\mu) = \frac{2\pi}{\beta_0\ln(\mu/\Lambda_{QCD})},
$$
where $\beta_0= 11-2n_f/3$ with $n_f$ being the number of quark flavors with
$m_q<\mu<m_{LQ}$.

Relation (\ref{a}) is universal one, since it is independent  
of whether the leptoquark is free or bound at
 the virtualities less than $\mu_0$.
In this work we include the evolution for the fragmentation into the heavy
lepto-quarkonium.

As one can see in Fig.\ref{a1-fig}, the leptoquark can lose about 20 \%
of its momentum before the hadronization.

\section{Integral probabilities of fragmentation}

\begin{figure}[t]
\hspace*{3cm}
\epsfxsize=9cm \epsfbox{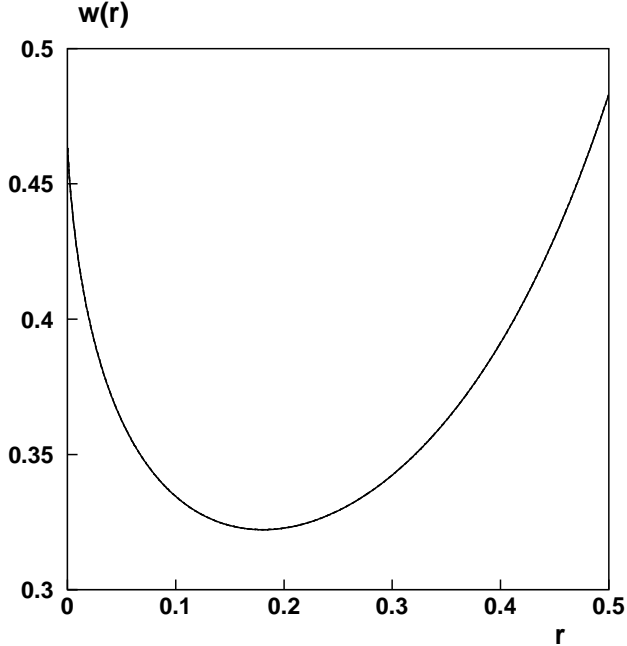}
\caption{The w-function for the leptoquark fragmentation into the heavy 
lepto-quarkonium versus the fraction of $r=m_Q/m$.}
\label{w-fig}
\end{figure}

As has been mentioned above the evolution conserves the integral probability
of fragmentation, which can be calculated explicitly from eq.(\ref{fra})
\begin{eqnarray}
\int dz\; D(z) &=& \frac{8\alpha_s^2}{27\pi}\; \frac{|R(0)|^2}{m_Q^3}\; 
{\rm w}(r),\\
{\rm w}(r) & = & \frac{[7+30r+20r^2+20r^3-75r^4-2r^5+
30r(1+r+3r^2+r^3)\ln r]}{15(1-r)^7}\; .
\end{eqnarray}
The function of ${\rm w}(r)$ is shown in Fig.\ref{w-fig} at low $r$.

To estimate numerically the yield of $(\bar b LQ)$ and $(\bar c LQ)$,
one has to evaluate the radial wave function at the origin from the 
non-relativistic potential models. As was found in \cite{Eichten,Ger},
the Martin and Buchm\" uller--Tye potentials give close results for the
heavy quarkonia with the reduced mass about $m_c\sim 1.5$ GeV. The
characteristics of charmed and beauty lepto-quarkonia are presented in Tab.1,
where we have used the Martin potential in the limit of infinitely heavy
leptoquark, $m\gg m_Q$, so that the reduced mass equals the heavy quark mass, 
and the level energy is given by the sum of quark mass and the binding
energy evaluated numerically from the Schr\" odinger equation.

\begin{table}[th]
\caption{The radial wave-functions at the origin, level energies and
average sizes of heavy lepto-quarkonia, as evaluated in Martin potential. }
\begin{center}
\begin{tabular}{|c|c|c|c|}
\hline
level & $R(0)$, GeV$^{3/2}$ & E, GeV & $\langle r\rangle$, fm \\
\hline
1S$(\bar c LQ)$ & 1.61 & 1.023 & 0.27\\
2S$(\bar c LQ)$ & 1.20 & 1.606 & 0.58\\
1S$(\bar b LQ)$ & 3.43 & 4.039 & 0.16\\
2S$(\bar b LQ)$ & 2.56 & 4.594 & 0.35\\
\hline
\end{tabular}
\end{center}
\end{table}

Then one finds for the probabilities of leptoquark fragmentation
into the 1S-states with $b$ and $c$-quarks: $P_w(b)=1.22\cdot 10^{-4}$ and
$P_w(c)=2.16\cdot 10^{-3}$, respectively, at $m_b=4.9$ GeV, $m_c=1.5$ GeV,
$\alpha_s(2m_b)=0.18$, $\alpha_s(2m_c)=0.26$ and $m=245$ GeV.

\begin{figure}[th]
\hspace*{3cm}
\epsfxsize=9cm \epsfbox{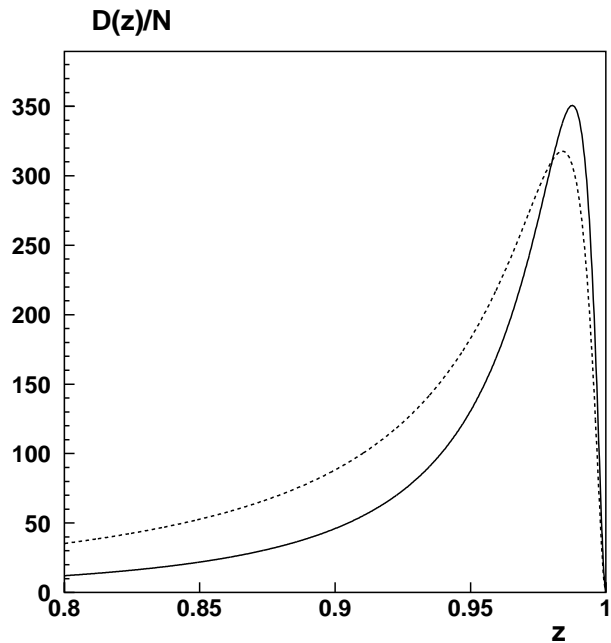}
\caption{The fragmentation function of leptoquark into the heavy 
lepto-quarkonium, the $N$-factor is determined by 
$N=\frac{8\alpha_s^2}{27\pi}\;
\frac{|R(0)|^2}{M^3 r^2(1-r)^2}$, the initial function -- solid line,
for the $(\bar b LQ)$-state with $r=0.02$,
the fragmentation function including the evolution -- dashed line, at
the scale $\mu$: $\frac{8\alpha_s}{3\pi}\ln \frac{\mu}{\mu_0}=0.25$.}
\label{d-fig}
\end{figure}

The perturbative fragmentation function in the leading $\alpha_s$-order
is shown in Fig.\ref{d-fig} at $r=0.02$. It is quite a hard distribution,
which becomes softer with the evolution (see Fig.\ref{d-fig}).

\section{Conclusion}

In this work the dominant mechanism for the production of possible 
bound states of a scalar leptoquark with a heavy anti-quark is considered for
high energy processes at large transverse momenta, where the fragmentation
contributes as the leading term.
The corresponding fragmentation function of scalar leptoquark into the
heavy lepto-quarkonium can be calculated in perturbative QCD, so that
for the S-wave states one finds
$$
D(z) = \frac{8\alpha_s^2}{27\pi}\;
\frac{|R(0)|^2}{M^3 r^2(1-r)^2}\;
\frac{z^2(1-z)^2}{(1-(1-r)z)^6}[(1+r^2)(1+(1-r)^2z^2)-2(1-r)^2(1+r)z],
$$
where $r$ is the ratio of heavy quark mass to the mass of the bound state.
In the infinitely heavy leptoquark limit, $D(z)$ has the form, which agrees
with what expected from the general consideration of $1/m$-expansion for
the fragmentation functions.

The hard gluon corrections caused by the splitting of scalar leptoquark
are taken into account so that the evolution kernel has the form
$$
P_{LQ\to LQ}(x,\mu) = \frac{4\alpha_s(\mu)}{3\pi}\;
\bigg[\frac{2x}{1-x}\biggr]_+,
$$
which results in the corresponding one-loop equations for the moments of
fragmentation function (see eqs.(\ref{da}), (\ref{a})).

The integral probabilities of scalar leptoquark fragmentation into the
charmed and beauty lepto-quarkonia are of the order of $10^{-3}$ and
$10^{-4}$, correspondingly.

\vspace*{4mm}
The author expresses his gratitude to profs. A.Wagner and P.M.Zerwas for a
kind hospitality and support during the visit to DESY, where this 
work was done, and to profs. S.S.Gershtein and A.K.Likhoded for
fruitful discussions and support.

This work is in part supported by the Russian Foundation for Basic Research,
grants 96-02-18216 and 96-15-96575.

\end{document}